\documentclass[conference]{IEEEtran}
\IEEEoverridecommandlockouts

\makeatletter
\def\markboth#1#2{\def\leftmark{\@IEEEcompsoconly{\sffamily}\MakeUppercase{\protect#1}}%
\def\rightmark{\@IEEEcompsoconly{\sffamily}\MakeUppercase{\protect#2}}}
\makeatother

\usepackage[english]{babel}
\usepackage{amsmath}
\usepackage{amsfonts}
\usepackage{amssymb}
\usepackage{amsthm}
\usepackage{array}
\usepackage{verbatim}
\usepackage{listings}

\usepackage{algorithm}
\usepackage{algpseudocode}
\usepackage{url}
\usepackage{enumerate}
\usepackage{multirow}

\usepackage{epsfig}
\usepackage{epstopdf}
\usepackage{graphicx}
\usepackage{caption}
\usepackage{subcaption}
\usepackage{multicol}
\usepackage[font=footnotesize]{caption}

\DeclareMathOperator*{\argmax}{arg\,max}

\def\beq{\begin{equation}}
\def\eeq{\end{equation}}
\def\beqa{\begin{eqnarray}}
\def\eeqa{\end{eqnarray}}
\def\beqan{\begin{eqnarray*}}
\def\eeqan{\end{eqnarray*}}

\title{An MDP Model for Optimal Handover Decisions \\ in mmWave Cellular Networks}

\author{{{\bf Marco Mezzavilla}$^*$, {\textbf{Sanjay Goyal}$^*$}, {\bf Shivendra Panwar}$^*$, {\bf Sundeep Rangan}$^*$, {\bf Michele Zorzi}$^\dagger$ }\\

$^*$ NYU Tandon School of Engineering, Brooklyn, NY, USA $^\dagger$ University of Padova, Italy\qquad\qquad   \\
\small{$\{$\texttt{mezzavilla,sanjay.goyal,panwar,srangan}$\}$\texttt{@nyu.edu}, \texttt{zorzi}\texttt{@dei.unipd.it}
}
}

\begin{document}
\maketitle

%


\begin{abstract}
The new frontier in cellular networks is harnessing the enormous spectrum
available at millimeter wave (mmWave) frequencies above 28~GHz. The challenging
radio propagation characteristics at these frequencies, and the use of highly
directional beamforming, lead to intermittent links between the base
station (BS) and the user equipment (UE). In this paper, we revisit the problem of
cell selection to maintain an acceptable level of service, despite the
underlying intermittent link connectivity typical of mmWave links.
We propose a Markov Decision Process (MDP) framework to study the properties and
performance of our proposed cell selection strategy, which jointly considers several
factors such as dynamic channel load and link quality.  We use the Value
Iteration Algorithm (VIA) to solve the MDP, and obtain the optimal set of
associations. We address the multi user problem through a distributed iterative approach, in which each UE characterizes the evolution of the system based on stationary channel distribution and cell selection statistics of other UEs. 
Through simulation results, we show that our proposed technique makes judicious handoff choices, thereby
providing a significant improvement in the overall network capacity.  Further,
our technique reduces the total number of handoffs, thus lowering the
signaling overhead, while providing a higher quality of service to the UEs.
\end{abstract}


\section{Introduction}

Cell selection is a fundamental functionality in wireless networks, and involves
choosing which base station (BS) the user equipment (UE) should be connected to.
Current cellular networks that operate in the microwave bands (around 2 GHz) use simple
heuristics to perform cell selection, usually choosing the BS that provides the
highest long-term signal to noise ratio (SNR) \cite{dahlman20134g}.

In this paper, we revisit the cell selection problem in the context of
next-generation cellular networks. These networks are expected to use
millimeter wave (mmWave) technology, that operates at frequencies above 28~GHz, 
thereby exploiting the enormous amount
of spectrum available in these bands.  At these frequencies, the radio
propagation characteristics are starkly different from their microwave counterparts.  
First, according to the Friis transmission
equation~\cite{Rappaport:Book}, the path loss can easily exhibit 30-40~dB more
attenuation. This higher path loss necessitates the
use of fairly narrow and very directional beams, that can be realized through phased antenna
arrays, whose implementation is made possible thanks to the smaller wavelengths 
that correspond to these 
frequencies.  Furthermore, due to the exacerbated blockage and shadowing effects~\cite{shadowing},
the wireless links will exhibit rapid variations in quality, thereby leading to
severe intermittency in link connectivity between the UE and the BS.

To address these challenges, and in particular to
maintain an acceptable level of service despite this intermittency, 
the density of BSs in mmWave cellular networks is expected to be an order
of magnitude higher than in current systems~\cite{bsdensity}. The UEs will track several BSs
simultaneously and rapidly switch between them in response to the fast-varying
link qualities~\cite{michele:mac}.  A simple approach to cell selection would
be for each UE to greedily pick the BS that is instantaneously the best, 
thereby attempting to keep an optimal
state for itself. Unfortunately, these approaches may lead to degraded performance
for other UEs~\cite{mitrl, andrews:load}, or even to instability. In addition, 
they entail significant overhead because of 
frequent signaling in the control plane due to the large number of BS handovers that would result under such a policy. 
Therefore, a better approach would need to consider the network behavior
and to look for solutions where all relevant information (including 
channel conditions and BS load) is explicitly included in the optimization.

The problem of cell selection in mmWave networks, as well as in macrocellular networks
with high mobility~\cite{vtc:train}, has received considerable attention over
the past few years. In~\cite{nsn:amitava}, Talukdar et al. conclude that in mmWave
the UE will remain associated with a BS for just a few seconds and, in some
cases, for as little as 0.75 s. In~\cite{michele:mac}, Shokri-Ghadikolaei et al. study the
implications of the mmWave PHY on the MAC layer and argue that, if simple
cell selection techniques based on SNR were used, the handovers would become
too frequent. Further, loss of channel information and outdated beamforming
vectors will also lead to more frequent outages and expensive cell discovery
searches. The impact of network load on cell selection in dense pico-cell
environments was studied by Ye et al. in~\cite{andrews:load}.  They argue
that considering SINR alone leads to sub-optimal assignments, and the optimal
approach is therefore to solve cell selection and resource allocation jointly.

One of the most popular techniques to study the problem of cell selection as
well as possible handoff strategies is through Markov Decision Processes
(MDPs)~\cite{mdp1, mdp2, mdp3}, which provide a useful mathematical framework
for studying the properties and performance of proposed cell selection
strategies. However, in order to be useful, MDPs need to be carefully applied.
In particular, it is important that all factors that play a key role in 
determining the goodness of a cell selection solution be included in the model.

Previous studies that only use partial information may lead to sub-optimal results. 
For example, Dang et al. in~\cite{mdp2} and Pan et al. in~\cite{mdp3} do
not consider dynamic variations in the network load to be an input to the cell
selection algorithm. Furthermore,~\cite{mdp2} considers a network with just one
UE, and therefore does not capture the global network-level performance effects of
the proposed technique. In~\cite{mdp1} Stevens-Navarro et al. consider a network with
just one BS, but with multiple relay nodes. 

A more comprehensive study of the cell association problem will need to (i) consider a network with multiple BSs, where the network load can vary dynamically, and (ii) explicitly include in the optimization problem the key parameters that affect the performance (at both the network and the user level) in a multi-cell multi-user scenario, including cell load and channel conditions. Our contribution in this paper is to develop 
such an approach and compare its performance to schemes that use only partial information.

The paper is organized as follows. In Sec. \ref{sec:model}, we introduce our model formulation. In Sec. \ref{sec:mdp}, we describe our decision algorithm and the iterative process applied to solve the multi-agent nature of the problem. In Sec. \ref{sec:results}, we discuss some simulation results. In Sec. \ref{sec:states}, we analytically derive the number of states for varying configurations. Finally, we conclude the paper and propose some future work in Sec. \ref{sec:conclu}.

\section{Model formulation}
\label{sec:model}
In our cell association problem, each UE can connect to a set of $L$ surrounding BSs. The time evolution of the quality of each link is described by a Markov process with $K$ states. $N$ represents the total number of UEs in the system. 
As a first step towards more general scenarios, we consider the situation in which 
all links have the same statistics. This can be
justified in mmWave scenarios where the channel can be assumed to alternate between well-defined states
(e.g., line-of-sight (LOS) and non-LOS) and all LOS links (non-LOS links) can be considered to be 
equally good (equally bad) on average. More general models where different statistics may be associated
to different links will be left for future work. 

Although the framework is general, we will illustrate the methodology using
a simplified version of the mmWave channel model described in \cite{mustafa}.
In this model, each link is characterized by three possible states:
\begin{itemize}
\item \emph{outage}, where no mmWave link is available;
\item \emph{LOS}, where a direct LOS mmWave link is available; 
\item \emph{NLOS}, where only a non-LOS mmWave link is available.
\end{itemize}

The presence of the outage state, which occurs due to blockage, and the highly dynamic behavior of the channel, which can move in and out of the outage state on a very short time scale, are unique to the mmWave model. In practice, the rate that a mobile experiences in
any state depends on several conditions, including interference and SNR.
However, to simplify the study, we will assume that the rate is uniquely determined
by the state. A more complex model, which includes the SNR or other variations within each
state, can also be included. The transmission rates in the LOS and NLOS states are based approximately on the average spectral efficiencies
in those states, as presented in \cite{mustafa}.

More specifically, the statistical model provided in \cite{mustafa} gives the probability that a UE is in each of the three states based on its distance from the BS. Assuming the UE is randomly dropped in each cell with a radius of 200 m (a typical cell radius in the mmWave range), we computed the steady-state probability of each state, $\pi$. Hence, via quadratic programming optimization, we obtain matrix $\mathbf{P}$ such that $diag(\mathbf{P})$ is close to $1-\frac{1}{T_{avg}}$, where $T_{avg}=[t_{LOS},t_{NLOS},t_{out}]$ is the average time spent in each state before leaving it. Also, this matrix is consistent with the steady-state equation, $\pi P = \pi$.  

Following this approach, we model the channel conditions seen by each UE towards the $L$ BSs as $NL$ i.i.d. Markov processes with common transition probability matrix,
\beq
\mathbf{P}=\begin{bmatrix}
p_{out-out} & p_{out-NLOS} & p_{out-LOS} \\ 
p_{NLOS-out} & p_{NLOS-NLOS}  & p_{NLOS-LOS} \\ 
p_{LOS-out} & p_{LOS-NLOS} & p_{LOS-LOS}
\end{bmatrix}.
\eeq

Each connection is defined as 
\beq
C^i=\{K_i,U_i\},
\eeq
where $K_i$ and $U_i$ represent the $K-$quantized channel state characterizing the link between a generic UE, $k$, and BS $i$ and the number of UEs connected to BS $i$, respectively. 

The state space is defined as a subset of
\beq 
S=[C^1 \times \mathbf{C^{L-1}]},
\eeq
constrained by \beq
\sum_{i=1}^{L}U_i=N,
\eeq
and where
\beq 
\mathbf{C^{L-1}} = [C^2 \times \ldots \times C^L].
\eeq

In a state, $C^1$ describes the primary connection, i.e., the BS serving UE $k$, and $\mathbf{C^{L-1}}$ represents the characterization of the $L-1$ surrounding links (between UE $k$ and the non-serving BSs). The state space $S$ contains all possible combinations of channel conditions towards the different BSs and load occupancy at each BS. Due to the symmetry of the system model, a single state can represent multiple situations (according to all possible permutations of a given scenario), which leads to a significant reduction of the number of states needed to represent the possible system configurations, and therefore to a better scalability of the model. To assess the complexity of these models, we analytically derive the number of those states in Section \ref{sec:states}.

%
%

The action is defined as the identifier of the cell that the UE will join at the next step, i.e.,

\beq \label{eq:actions}
a \in A_s, \quad A_s=\{1, \ldots, L\},
\eeq
which corresponds to a handover if $a$ is different from the current serving cell. Here, $A_s$ represents the set of all BSs.


In an MDP, the statistics of the next state depends only on the current state and on the decision made. Therefore, we need to define the transition probabilities, $p(s_j|s_i,a)$, i.e., the probabilities of moving to state $s_j$ given the current state $s_i$ and under action $a$, where $s_i, s_j \in S$. 
The transition probabilities must satisfy the condition
\beq
\sum_{s_j\in S}p (s_j|s_i,a)=1, \forall s_i \in S, a \in A_s.
\eeq

\begin{algorithm}[t!]
\caption{MDP-based Handover}\label{euclid}
\begin{algorithmic}[1]
\State Intialization: Initial Policy $\mathcal{D}_{0}^{i}, \forall i \in [1,N]$
\State For each iteration ($n > 0$)

\hspace{-0.8cm}
Select a UE: $k = mod(N, n) + 1$

\hspace{-0.8cm}
Update the Policies:

For UE $k$

~~~~$p(s_j|s_i,a) = f(\mathcal{D}_{n-1}^{-\mathbf{k}}, r(s_i,s_j,a))$ 

~~~~~~~~~~~~~~~~~~~~~~~~~~~$\forall s_i,s_j \in S, \forall a$

~~~~$\mathcal{D}_{n}^{k} =$ VIA$(\bold{p}, \bold{r})$

For other UEs $m\in \{1,,2,\dots,N\}\setminus{k}$

~~~~$\mathcal{D}_{n}^{m} = \mathcal{D}_{n-1}^{m}$

\hspace{-0.8cm}
$n$ = $n+1$

\hspace{-0.8cm}
Until convergence   
 \end{algorithmic}
\label{algo1}
\end{algorithm}

Our link reward function $r(s_i,a)$ depends on the average reward over all possible destinations. Thus, we let $r_t(s_i,a,s_j)$  denote the value at time $t$ of the instantaneous reward received given that the state of the system at decision epoch $t$ is $s_i$, action $a \in A_s$ is selected, and the system is in state $s_j$ at decision epoch $t+1$. Its expected value at decision epoch $t$ can be evaluated by computing 
\beq
r(s_i,a)=\sum_{s_j\in S}r_t(s_i,a,s_j) p_t(s_j|s_i,a),
\eeq
where 
\beq
r_t(s_i,a,s_j)=(1-c(s_i,s_j,a)) \frac{R_{s_j}}{U_{a}+1},
\eeq
$R_{s_j}$ is the achievable rate the UE would enjoy if it were the only UE in its cell, and only depends on the channel quality of state $s_j$, whereas $\frac{R_{s_j}}{U_{a}+1}$ is the rate actually available when the cell load $U_{a}$ on the target BS $a$ is taken into account.\footnote{Note that in this formulation the cell load does not include the incoming user (which instead accounts for the "+1" in the denominator). Also, $U_a+1$ is an estimate based on the status of the BS occupancy in the previous slot, and does not necessarily represent the true state, which depends on the decisions being made in the current slot (i.e., other users leaving or joining that BS).} $c(s_i,s_ja)$ is the handover cost function, which is equal to $OH$ if the UE moves from the associated BS in state $s_i$ to a different BS in state $s_j$, and is otherwise equal to zero. We define the value of $OH$ as a percentage of the spectrum that needs to be used for signaling.



\section{MDP-based handover decision}
\label{sec:mdp}
In this section, we describe our algorithm to obtain the optimal cell selection strategy for each UE. We use a distributed iterative approach in which each UE finds its optimal deterministic policy when assuming that all other UEs make handover decisions based on current BS occupancy but assuming steady-state channel conditions, which results in an approximated cell occupancy evolution\footnote{We note that a precise model would need to keep track of the channel conditions from all UEs to all BSs, which is clearly an impossible task.}.

The proposed algorithm is described in Algorithm~\ref{algo1}. It initializes the system with a random policy assignment to each UE, $\mathcal{D}_{0}^{i}, \forall i \in \{1,2,\dots,N\}$, where $\mathcal{D}_{n}^{i} = (d^n_i(s_1),d^n_i(s_2),\cdots,d^n_i(s_j),\cdots)$ contains the set of actions for all states $s_j \in S$ for UE $i$ at iteration $n$. This algorithm runs for multiple iterations. In each iteration, a UE $k$ is selected sequentially among the given set of UEs. For the selected UE, the policy is updated based on the policies of the other UEs at the previous iteration, denoted by $\mathcal{D}_{n-1}^{-\mathbf{k}}$. We introduce the following definitions.

\beq
s(t)=[(K_1(t),U_1(t)), \cdots, (K_L(t),U_L(t))]
\eeq
is the state seen by the selected UE, i.e., the UE that is updating its policy. On the other hand,
\beq
s^{\star}(t)=[(.,U_1(t)), \cdots, (.,U_L(t))],
\eeq
is the approximate description of the state associated with all the other UEs, which refers to the fact that for those UEs we define a decision strategy that only refers to the cell occupancy, thereby avoiding the need to track instantaneous channel conditions for everyone. That being said, we can define the probability for any UE (say UE $x$) to select BS~$i$ starting from any state $s^{\star}(t)$ as

\begin{algorithm}[t!]
\caption{Value Iteration Algorithm (VIA($\bold{p}, \bold{r}$))}\label{euclid}
\begin{algorithmic}[1]
\State Select $v^0 \in V$, specify $\epsilon, \omega> 0$, and set $n=0$
\State For each $s_i \in S$, compute $v^{(n+1)}(s_i)$ by

\hspace{-0.8cm}
{\footnotesize $v^{(n+1)}(s_i)=\max_{a \in A_s}\left \{ r(s_i,a)+\sum_{s_j \in S} \omega~p(s_j|s_i,a)~v^{n}(j) \right \}$}
\If {$\begin{Vmatrix}
v^{n+1}-v^n
\end{Vmatrix}<\epsilon (1- \omega)/2 \omega$}

\textbf{goto} step $\mathbf{6}$
 \Else 
 
$n \gets n+1$

\textbf{goto} step $\mathbf{2}$
\EndIf\\
For each $s_i$, choose

\hspace{-0.8cm}
{\footnotesize $d_k(s_i) = \argmax_{a \in A_s} \left \{ r(s_i,a)+\sum_{s_j \in S} \omega~p(s_j|s_i,a)~v^{n}(j) \right \}$}
\end{algorithmic}
\label{algo2}
\end{algorithm}

\beq
P[s^{\star}(t) \rightarrow i]=  \sum_{m=1}^{K^L}\pi(s_{m}(t))~\delta_{id_{x}(s_m(t))},
\eeq
where $K^L$ is the total number of channel states' combinations, and $\delta_{id_{x}(s_m(t))}$ is calculated from the policy of the corresponding UE. 
\beq
\delta_{id_x(s_m(t))}=\left\{\begin{matrix}
1,  \quad i = d_{x}(s_m(t))\\ 
0,  \quad  i \neq d_{x}(s_m(t))
\end{matrix}\right..
\eeq

\begin{figure}[t]
  \centering
    \includegraphics[trim = 0mm 0mm 0mm 7mm, clip, width=1\columnwidth]{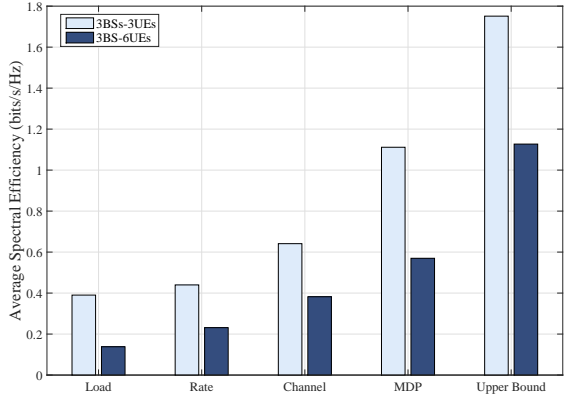}
      \caption{Average Spectral Efficiency (bits/s/Hz) for different schemes with 3 BSs, handover cost = 10\%.}
            \label{ase}
\end{figure} 

Here, $\pi(s_{m}(t))$ represents the steady-state distribution of the channel. Now, because we have introduced a probabilistic evolution of each UE, we can update the transition matrix accordingly. Then, based on the updated transition probability matrix ($\bold{P}$), the Value Iteration Algorithm (VIA) described in Algorithm \ref{algo2} is used to solve the MDP, which gives the optimal deterministic policy of the selected UE. This way, the load occupancy dynamics will be captured, thus allowing us to evaluate the overall performance of a fully characterized multi user system. In our VIA algorithm, $v(\mathbf{s})$ denotes the maximum expected total reward, $\omega$ represents the discount factor, i.e., the length of the analyzed horizon, whereas $\bold{r}$ is the vector containing the reward values. 
For other UEs, the policy is retained from the previous iteration. The reason behind solving the MDP for one UE per iteration is to reach convergence. If multiple UEs change their policy simultaneously, we observed that the algorithm does not converge and instead oscillates indefinitely. This is a very well known property of multi-user distributed solutions, where the performance of a user is strongly coupled with the performance of the other users. For example, this issue is discussed in the context of distributed power control in a multi-cell 4G network \cite{zhang2011weighted}. In this paper, we do not provide a formal proof of the convergence of the proposed algorithm, which is left for future work.

In summary, (i) we derive the optimal policy at each UE as a function of its detailed state (i.e., channel state plus occupancy), and find the related steady-state distribution; (ii) we converge to an optimal equilibrium through an iterative process by averaging the previous policies of all users over the conditional channel distribution at each iteration.

\begin{figure}[t]
  \centering
    \includegraphics[trim = 0mm 0mm 0mm 7mm, clip, width=1\columnwidth]{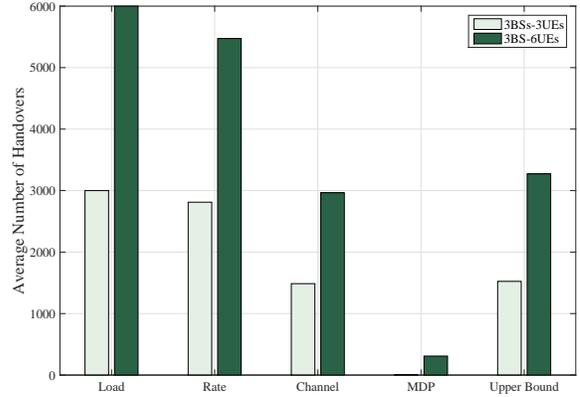}
      \caption{Number of Handovers for all different schemes with 3 BSs, handover cost = 10\%.}
      \label{hos}
\end{figure}

\section{Simulation results}
\label{sec:results}
In this section, we present some simulation results obtained by applying the proposed MDP-based handover algorithm for varying system configurations. Specifically, we studied simple scenarios where the number of BSs is fixed to $3$, while the number of UEs varies between $3$ and $6$. Moreover, to better assess the performance of the proposed model, we generate the results for different handover costs, defined as the percentage of resources spent for signaling and flow rerouting, as defined in Section~\ref{sec:model}. 
The channel matrix shown in (\ref{channel_matrix}) is obtained as per the description given in Section~\ref{sec:model}, where $T_{avg}=[t_{LOS},t_{NLOS},t_{out}]=[5,25,3]$, which characterizes an urban scenario where the dominant link is  NLOS:
\beq
\mathbf{P}=\begin{bmatrix}
0.55 & 0.3 & 0.15  \\ 
0.01 & 0.8  & 0.19  \\ 
0.38  & 0.40 & 0.22
\end{bmatrix}.
\label{channel_matrix}
\eeq

In addition, we consider a 28~GHz carrier frequency with 1~GHz bandwidth, a slot duration equal to 125 $\mu$s and $30$ OFDM symbols per slot ($6$ for control, $24$ for data). 

We use the optimal policy obtained with our algorithm, and compare its performance against other cell selection approaches, namely: \begin{itemize}
\item \textbf{Load:} Each UE connects to the least loaded BS. If two or more BSs show the same occupancy level, UEs randomly select one of them; 
\item \textbf{Rate:} UEs associate with the BS that can offer the best instantaneous rate, which depends on both channel and load information;
\item \textbf{Channel:} Traditional approach where UEs select the BS offering the best channel (SNR-based);
\item \textbf{Upper Bound:} Centralized exhaustive search method; it requires global information about link qualities along with cell occupancy, and exploits UE coordination, which is unavailable in distributed schemes. Hence, this approach represents an upper bound.
\end{itemize}

In Figs. \ref{ase} and \ref{hos}, we report the results for the case of $3$ UEs with $3$ BSs and handover cost 10\%. We plot the average spectral efficiency (bits/s/Hz) and the average number of handovers. We can observe how the optimal policy obtained by solving the MDP described in Section \ref{sec:mdp} outperforms other approaches. In particular, we can note that the \emph{Load}-based scheme, which relies solely on occupancy information, is very inefficient and results in biasing all the UEs towards unloaded cells. As a consequence, BSs will be overloaded, thus explaining the low rate observed in Fig. \ref{ase}. On the other hand, \emph{Channel}- and \emph{Rate}-based schemes show better performance but, because of the channel variations that characterize mmWave links, instantaneous actions are highly inefficient.

Instead our MDP model, where the dynamics of the links are fully captured, can be seen to provide significantly better performance. This not only results in increased sum-rate, but also provides a greatly reduced number of handovers, as shown in Fig. \ref{hos}, thus representing a more energy-efficient solution. The \emph{Upper Bound} refers to a centralized scheme, which compared to our distributed scheme has the advantage of full knowledge and of coordinated decisions, thereby resulting in significantly better performance in general. Nevertheless, it can be observed that despite such big advantages the performance gap between our solution and the centralized upper bound is not very wide, showing that our solution (which is not necessarily the distributed optimum because the problem is non-convex) still achieves a fairly good performance

\begin {table}
\caption {Average Spectral Efficiency Gain (bits/s/Hz)} \label{tab:tab3} 
\label{results}
\begin{center}
    \begin{tabular}{| l | l | l | l | l | l |}
    	\hline
		 & \textbf{3\% OH} & \textbf{6\% OH} & \textbf{10\% OH}  & \textbf{30\% OH}  \\ \hline
		\textbf{3 UEs} & 37\% & 39\% & 42\% & 51\% \\ \hline
		\textbf{4 UEs} & 33\% & 35\% & 39\% & 50\% \\ \hline
		\textbf{5 UEs}  & 30\% & 32\% & 36\% & 50\% \\ \hline
		\textbf{6 UEs}  & 26\% & 28\% & 33\% & 50\% \\ \hline		
    \end{tabular}
\end{center}
\end{table} 


In Table \ref{results}, we report a more detailed sum-rate comparison of our \emph{MDP}-based model against a traditional \emph{Channel}-based association scheme in terms of average spectral efficiency gain (\%). We can observe significant gains in the \emph{MDP}-based approach, which increase as the HO cost increases. The ability to capture optimal solutions for more complex scenarios may lead us to draw some important conclusions about the shape of effective policies, and to find heuristics to better address a number of critical issues related to mmWave cellular networks.

\section{Complexity analysis}
\label{sec:states}


In this section, we aim at analyzing the complexity of our MDP model in terms of number of states required as a function of the number of UEs and BSs.

First of all, we will analytically derive the number of load occupancy combinations, for at most $5$ BSs,\footnote{A reasonable maximum number of surrounding BSs.}.
Let us introduce $q(\cdot)$, expressed as 

\beq
q(x)=\max\{x,0\}.
\eeq

Now, we can count the possible occupancy combinations for $1, 2, 3, 4$ BSs as a function of the number of UEs $N$, i.e.,

\beq
c_{1}(N)=N+1 ,
\eeq

\beq
c_{2}(N)=\left \lfloor \frac{N}{2} \right \rfloor,
\eeq

\beq
c_{3}(N)=\sum_{i=0}^{N-1} q\left ( \left \lfloor \frac{N-1-i}{2} \right \rfloor - i  \right ),
\eeq

\beq
\begin{split}
c_{4}(N) &= \sum_{i=0}^{N-1}\sum_{j=0}^{N-1} q\left ( \left \lfloor \frac{N-1-j}{2} \right \rfloor - j\right ) - \\
           &   - \sum_{k=0}^{i-1} q\left ( \left \lfloor \frac{N-2-i-k}{2} \right \rfloor - k\right ),
\end{split}
\eeq
and derive the number of load occupancy combinations at varying number of BSs, $L$, as follows, 

\beq
C_L(N)=c_1(N)+\sum_{i=0}^{N-1}\sum_{k=2}^{L-1} c_k(N-i).
\eeq

\begin{figure}[t!]
  \centering
    \includegraphics[trim = 0mm 0mm 15mm 0mm, clip, width=1\columnwidth]{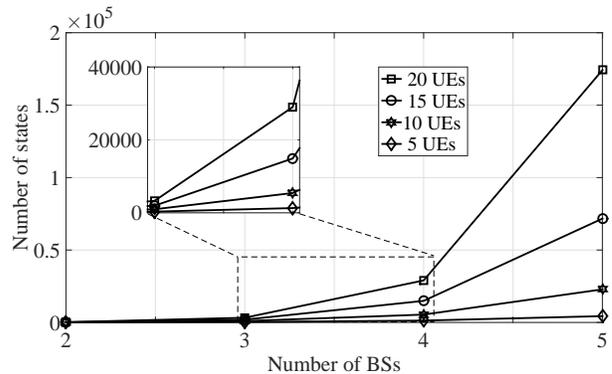}
      \caption{Number of states at varying number of BSs, $L$, and UEs, $N$.}
      \label{ues}
\end{figure}

As stated above the number of possible channel states' combinations is equal to $K^L$.
Therefore, the total number of states, as shown in Fig. \ref{ues}, will be as follows:
\beq
|S(L,K,N)|=K^LC_L(N).
\eeq
 
In terms of run time, the computational complexity (convergence time) of the proposed MDP approach increases exponentially with
the number of UEs. However, the proposed algorithm can be executed offline within specific clusters of mmWave cells, at varying number of associated users. 
Each BS will disseminate the optimal policies for various numbers of instantaneous connected UEs, thus quickly adapting to topology changes, i.e., a new UE coming or leaving.

\section{Conclusions}
\label{sec:conclu}
In this paper, we have argued why harnessing the potential of mmWave cellular
networks requires revisiting the problem of cell selection. Judicious choices
in cell selection serve to improve the quality of service and increase network
capacity, while minimizing the signaling overhead caused by sub-optimal cell
selections and subsequent re-associations.  We have made the case in favor of
using MDPs to design and analyze association techniques.  Through numerical
analysis and simulations, we have demonstrated the ability of our proposed
technique to achieve these goals. In the future, we plan to extend this work in several ways: $i)$ evaluate over more complex networks to derive procedural guidelines to design heuristics; $ii)$ investigate whether finer-grained SNR measurements can improve
outcomes; $iii)$ examine the effectiveness of these techniques in heterogeneous
networks. In conclusion, although mmWave technology holds the promise to revolutionize
cellular networks, realizing this potential will
require revisiting and potentially redesigning several components of the
communication stack. This paper makes an important step in this direction with focus on the problem of cell selection in mmWave cellular networks.

\bibliographystyle{IEEEtran}
\bibliography{biblio.bib}

\end{document}